\documentclass[aps,prl,comment-prl,twocolumn,groupedaddress,amssymb,showpacs]{revtex4}
\begin{document}

\preprint{APS/123-QED}
\title{Comment on ``Clock Shift in High Field Magnetic Resonance of Atomic Hydrogen''}
\author{A.\,I. Safonov$^1$}
\email{safonov@isssph.kiae.ru}
\author{I.\,I. Safonova$^1$}
\author{I.\,S. Yasnikov$^2$}
\affiliation{
$^1$Russian Research Centre Kurchatov Institute, pl. Kurchatova 1, Moscow, 123182 Russia\\
$^2$Togliatti State University, Togliatti, 445667, Russia}

\date{\today}


\maketitle

Ahokas {\it et al.}~\cite{Ahokas_3D} measured the hyperfine frequency shifts in three-dimensional
spin-polarized atomic hydrogen by means of ESR. In this Comment, we address their analysis of the
interaction energy of the ground-state H atoms in different hyperfine states and show that the
quoted difference $\Delta a=a_t-a_s$ between the triplet and singlet scattering lengths derived
from the correctly measured shifts is overestimated by a factor of two.

Ahokas {\it et al.} observed the transitions $a\rightarrow d$ and $b\rightarrow c$ in the presence
of the third-state atoms ($b$ and $a$, respectively) and found the shifts of the corresponding
resonance fields to be $\Delta B_{ad}=C_{ab}n_b+C_{aa}n_a$ and $\Delta B_{bc}=C_{bb}n_b+C_{ba}n_a$
with $C_{ab}\approx C_{ba}=8(2)\times10^{-19}$~cm$^3$ and $C_{aa}$, $C_{bb}\ll C_{ba}$. To explain
this observation, Ahokas {\it et al.} expressed the spin states of a pair of atoms in the
$|S,m_S,I,m_I\rangle$ basis, that is, in terms of electron and nuclear singlets and triplets. In
particular, in the high-field limit,
$|ab\rangle=\frac{1}{\sqrt{2}}(|e_t,n_t\rangle+|e_t,n_s\rangle)$,
$|ac\rangle=|bd\rangle=\frac{1}{2}(|e_t,n_t\rangle+|e_t,n_s\rangle+|e_s,n_t\rangle+|e_s,n_s\rangle)$.
At low temperature, only the symmetric components $|e_t,n_t\rangle$ and $|e_s,n_s\rangle$
contribute to ($s$-wave) scattering via the triplet $V_t$ and singlet $V_s$ interatomic potential,
respectively.

More specifically, the interaction energy of H atoms in different hyperfine states $\alpha$ and
$\beta$ ($\alpha, \beta=a,b,c,d$) is given by the second-quantization Hamiltonian
($i,j=\alpha,\beta$; $i\neq j$)~\cite{Dau9}
\begin{equation}
\hat{H}^{\rm int}_{\alpha\beta} = \frac{1}{2}\sum_{{\bf k}{\bf q}{\bf k}'{\bf q}'ij}\langle {\bf
k}'i,{\bf q}'j|V|{\bf k}i,{\bf q}j\rangle \hat{a}^+_{{\bf k}'i}\hat{a}^+_{{\bf q}'j}\hat{a}_{{\bf
q}j}\hat{a}_{{\bf k}i}.
\end{equation}
The wavefunction of two bosons must be symmetric,
\begin{eqnarray}
|{\bf k}i,{\bf q}j\rangle = \frac{1}{\sqrt{2}}(R_+|ij\rangle_+ + R_-|ij\rangle_-),
\end{eqnarray}
where the spatial and spin parts are, respectively, $R_\pm=\frac{1}{\sqrt{2}}(\psi_{\bf k}({\bf
r}_1)\psi_{\bf q}({\bf r}_2)\pm\psi_{\bf q}({\bf r}_1)\psi_{\bf k}({\bf r}_2))$ and
$|ij\rangle_\pm=\frac{1}{\sqrt{2}}(|ij\rangle\pm|ji\rangle)$. The use of the symmetric form (2) of
the diatomic wavefunction instead of simply $|{\bf k}i,{\bf q}j\rangle=\psi_{\bf k}({\bf
r}_1)\psi_{\bf q}({\bf r}_2)|ij\rangle$ does not change the sum (1) because the bosonic creation
(annihilation) operators $\hat{a}^+_{{\bf k}i}$ and $\hat{a}^+_{{\bf q}j}$ ($\hat{a}_{{\bf k}i}$
and $\hat{a}_{{\bf q}j}$) with $i\neq j$ obviously commute. The interaction strength $\lambda$ of
the pseudopotential $V({\bf r}_2-{\bf r}_1)=\lambda\delta({\bf r}_2-{\bf r}_1)$ has the
eigenvalues $\lambda_s$ or $\lambda_t$ corresponding to the spin states $|e_s\rangle$ or
$|e_t\rangle$ of the atomic pair. As such, the potential is nearly independent of nuclear spins:
$\langle e_tn_s|\lambda|e_tn_s\rangle=\langle e_tn_t|\lambda|e_tn_t\rangle=\lambda_t$ and $\langle
e_sn_t|\lambda|e_sn_t\rangle=\langle e_sn_s|\lambda|e_sn_s\rangle=\lambda_s$ (here we write only
the {\it spin} parts of the matrix elements). Instead, Ahokas {\it et al.} used $\langle
e_tn_s|\lambda|e_tn_s\rangle=\langle e_sn_t|\lambda|e_sn_t\rangle=0$ arguing that the
antisymmetric states do not scatter via $s$-waves. This zeroing is only valid if the matrix
elements include the vanishing {\it spatial} factor $\langle R_-|\delta({\bf r}_2-{\bf
r}_1)|R_-\rangle$. Actually, it is this spatial factor, which cancels the contribution of the
antisymmetric states to s-wave scattering and to the interaction energy (1).
On the contrary, the spatial part of the matrix elements for the symmetric states
$R_+|ij\rangle_+$ is doubled. In other words, the atoms of a heterostate symmetric pair behave as
identical.

Clearly, $|ab\rangle_+=|e_tn_{t,0}\rangle$ and
$|ac\rangle_+=|bd\rangle_+=\frac{1}{\sqrt{2}}(|e_{t,0}n_{t,0}\rangle+|e_sn_s\rangle)$.
Consequently, $\lambda_{ab}^+\equiv\langle ab|\lambda|ab\rangle_+=\lambda_t$ and
$\lambda_{ac}^+=\lambda_{bd}^+=\frac{1}{2}(\lambda_s+\lambda_t)$. Then, according to Eqs. (1),
(2), the interaction energies are $E_{ab}=\lambda_tn_an_b$,
$E_{ac}=\frac{1}{2}(\lambda_s+\lambda_t)n_an_c$, $E_{bd}=\frac{1}{2}(\lambda_s+\lambda_t)n_bn_d$
and the shifts become
\begin{eqnarray}
\hbar\Delta\omega_{bc}=\frac{2\pi\hbar^2}{m}(a_s-a_t)n_a,\\
\hbar\Delta\omega_{ad}=\frac{2\pi\hbar^2}{m}(a_s-a_t)n_b,
\end{eqnarray}
in agreement with our previous result~\cite{Safonov_08}. Note, that Ahokas {\it et al.} consider
the atoms of a symmetric heterostate pair as distinguishable and therefore lose a factor of two in
the expressions for the shifts. From Eqs. (3) and (4) we find
$C_{ba}=C_{ab}=\frac{\hbar}{\gamma_em}(a_t-a_s)$ and $\Delta a=30(5)$~pm. This is somewhat lower
than the theoretical values $\Delta a=42-55$~pm for the difference in the scattering
lengths~\cite{Williams, Jamieson, Chakraborty}.


\begin{acknowledgments}
This work was supported by the Russian Foundation for Basic Research, project no. 09-02-01002-a.
\end{acknowledgments}

\bibliography{Comment_Ahokas}

\end{document}